\documentclass[english,aps, paper, twocolumn]{revtex4}
\usepackage[T1]{fontenc}
\usepackage[utf8]{inputenc}
\usepackage{color}
\usepackage{babel}
\usepackage{graphicx}
\usepackage[unicode=true,pdfusetitle,
 bookmarks=true,bookmarksnumbered=false,bookmarksopen=false,
 breaklinks=false,pdfborder={0 0 1},backref=false,colorlinks=false]
 {hyperref}

\makeatletter
\@ifundefined{textcolor}{}
{%
 \definecolor{BLACK}{gray}{0}
 \definecolor{WHITE}{gray}{1}
 \definecolor{RED}{rgb}{1,0,0}
 \definecolor{GREEN}{rgb}{0,1,0}
 \definecolor{BLUE}{rgb}{0,0,1}
 \definecolor{CYAN}{cmyk}{1,0,0,0}
 \definecolor{MAGENTA}{cmyk}{0,1,0,0}
 \definecolor{YELLOW}{cmyk}{0,0,1,0}
}

\usepackage{lmodern}  

\makeatother

\begin{document}

\title{Rayleigh scattering in coupled microcavities: Experiment}

\author{Z. Vörös$^{1}$, P. Mai$^{1}$, M. Sassermann$^{1}$, G. Weihs$^{1}$,
A.M. Andrews$^{2}$, H. Detz$^{2}$, G. Strasser$^{2}$}

\affiliation{$^{1}$Department of Experimental Physics, University of Innsbruck,
Technikerstraße 25/d, Innsbruck, A-6020 Austria}

\affiliation{$^{2}$Vienna University of Technology, Institute for Solid State
Electronics, Floragasse 7, A-1040 Vienna, Austria}
\begin{abstract}
We experimentally analyze Rayleigh scattering in coupled planar microcavities.
We show that the correlations of the disorder in the two cavities
lead to inter-branch scattering of polaritons, that would otherwise
be forbidden by symmetry. These longitudinal correlations can be inferred
from the strength of the inter-branch scattering.
\end{abstract}
\maketitle

\section{Introduction}

In the last two decades, microcavity polaritons have proved a valuable
testbed in the investigation of a number of interesting phenomena,
such as non-equilibrium condensation \cite{Kasprzak2006,Balili2007},
superfluidity \cite{Amo2009b,Cristofolini2013}, lasing \cite{Weihs2003,Nelsen2009},
parametric oscillations \cite{Savvidis2000a,Tartakovskii2002,Giacobino2005},
soliton formation \cite{Hivet2012,Sich2014} to name but a few. A
recent review of the experiments and related theory can be found in
\cite{Carusotto2013}. Polaritons emerge as the eigenstates of strongly
coupled quantum well excitons, and photons of a planar cavity of length
of the order of the wavelength of light. Given that the system possesses
two-dimensional translational invariance, both the photon, and exciton
modes can be described by the in-plane momentum, and consequently,
photons with a given momentum will couple to excitons with the same
momentum. It is also worth noting that in single cavities, the polariton
branches are not degenerate anywhere.

The structural disorder induced by the fabrication process results
in fluctuations of the exciton and photon potentials; the translational
symmetry is broken. This gives rise to coherent resonant Rayleigh
scattering (RRS) of the polaritons. While initially RRS was regarded
an unwanted feature, later it was realized that useful information
can be gained about the structural defects by studying the properties
of the scattered light \cite{Langbein2010}. It was pointed out as
well that the instantaneous and polarization-conserving scattered
light can be utilized in testing for superfluidity \cite{Carusotto2004,Christmann2012,Amo2009b}.

In a single cavity, given the isotropic polariton dispersion relations,
the resonant nature of the scattering leads to an annular RRS emission
in the far field \cite{Freixanet1999,Houdre2000,Langbein2002}. The
wave vectors of the final states populating the emission ring have
the same length. 

Rayleigh scattering has been well known from experiments with quantum
wells (QW), where it was used to probe potential fluctuations \cite{Langbein1999}.
As a consequence, in early theoretical studies of RRS of polaritons,
it was assumed that the disorder in the quantum wells is the main
cause \cite{Whittaker2000,Shchegrov2000}. 

Several experimental results, however, gave strong indication that
the dominant contribution comes from the photonic part of the polaritons.
Gurioli et al., e.g., observed RRS at large negative detuning, where
the polariton is predominantly photonic \cite{Gurioli2001}. More
recently, Maragkou et al. demonstrated RRS in a purely photonic cavity
\cite{Maragkou2011}. In several experiments, it has also been found
that RRS produces a cross-shaped pattern in addition to the well-known
ring. This feature was explained by a growth-induced cross-hatch pattern
in the cavity mirrors, underlining once more that RRS is caused by
the photonic component of polaritons \cite{Gurioli2001,Langbein2002,Langbein2004,Zajac2012c}.

It is not clear a priori, what the emission pattern will be, if polariton
branches can be arranged in a way that they are degenerate: in such
a case, the resonance condition could be satisfied by states that
might belong to different symmetry classes, leading to multiple concentric
rings in the far field. In particular, in a double cavity, due to
the reflection symmetry with respect to the mirror between the two
cavities, these branches possess, in increasing order of the energy,
even, odd, even, and odd parity, and as a consequence, scattering
would be allowed only from the first to third, and second to fourth
branches, but these pairs are not degenerate. 

Recently, inter-branch Rayleigh scattering has been observed in a
triple cavity where the splitting of the three cavity modes and strong
exciton-photon coupling at the same time gives rise to six polariton
branches \cite{Abbarchi2012}. The experimental findings were interpreted
using structural data gained by scanning electron microscope and X-ray
diffraction measurements. In detail, it was demonstrated that a periodic
modulation induced by growth manifests itself in scattering into preferred
directions in momentum space. However, in these experiments, polaritons
were injected with zero momentum, and this configuration does not
allow for the simultaneous study of inter- and intra-branch scattering.

In this paper, we will study polariton RRS in a coupled double cavity.
We will focus on an experimental configuration where polaritons are
injected at an in-plane momentum that allows observation of inter-
and intra-branch scattering at the same time. 

A simple model, based on the extension of the theory introduced by
Savona \cite{Savona2007}, has been developed to explain the breaking
of the parity symmetry in our system \cite{Voros2014}.

\section{Experiment}

\begin{figure}[h]
\includegraphics[width=0.4\columnwidth]{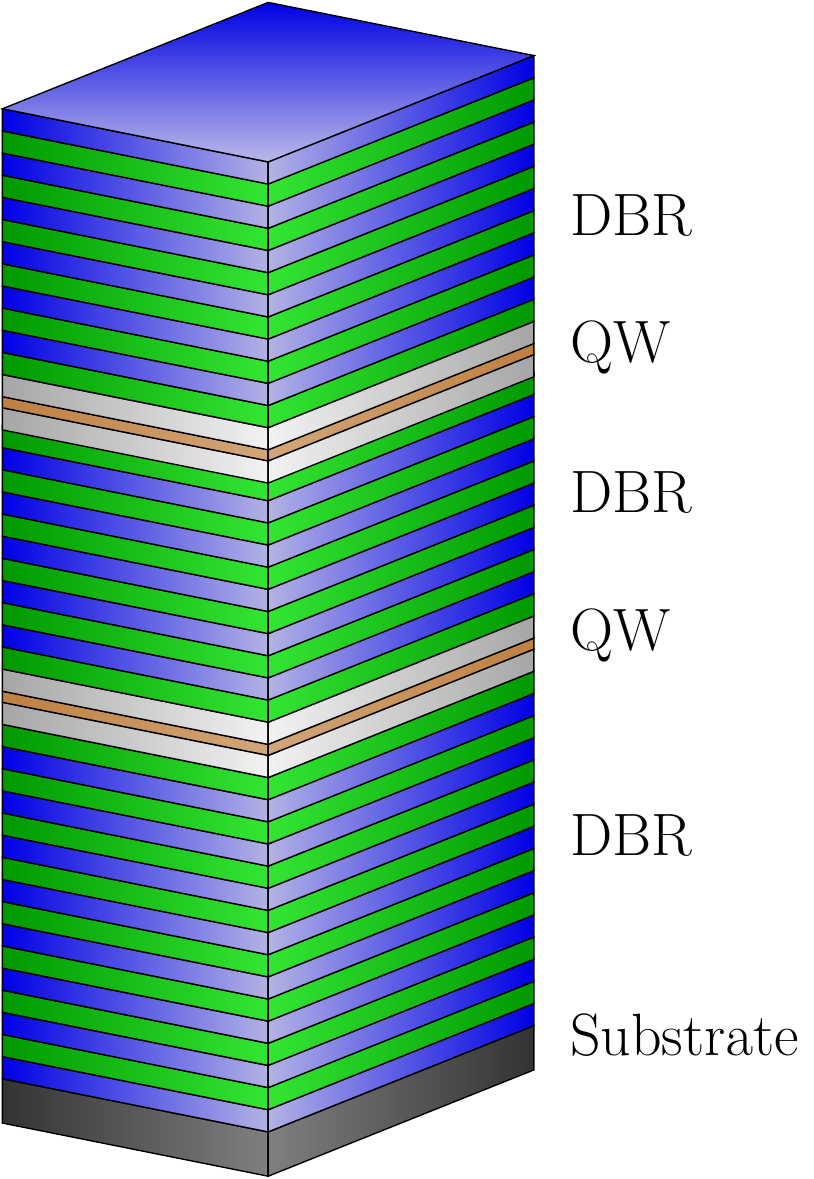}\includegraphics[width=0.6\columnwidth]{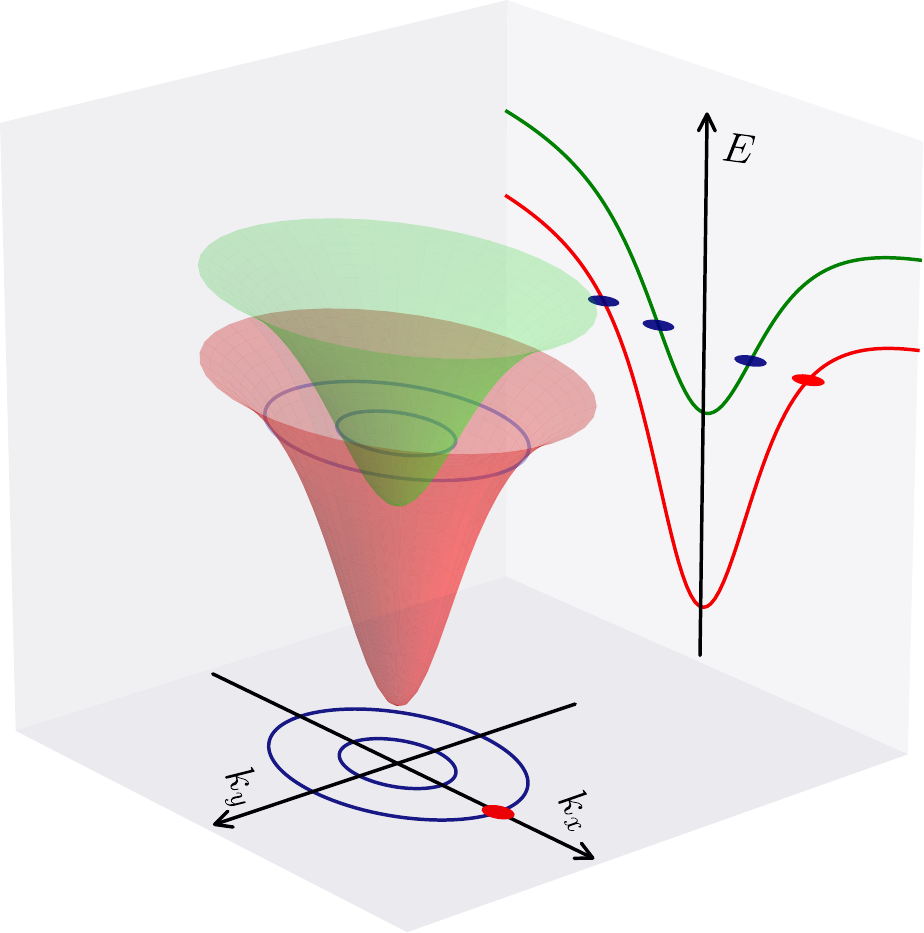}\caption{Left: Schematic drawing of the double cavity structure used in the
experiments. Right: polariton dispersions, and the scattering process
considered in the paper. Only the lower two polaritons are plotted.
The red branch is symmetric, while the green is anti-symmetric. Also
shown are the resonant Rayleigh rings in blue. The red dot is the
pump.}

\label{fig:doublecavity}
\end{figure}

The double cavity system, grown by molecular beam epitaxy, can be
seen on the left hand side of Fig.\,\ref{fig:doublecavity}. Two
$\lambda/2$ cavities each containing 4 70-$\mathrm{\AA}$ GaAs quantum
wells are coupled via an intermediate distributed Bragg reflector
(DBR). The top and bottom DBRs consist of 16 periods of $\mathrm{Al}{}_{0.2}\mathrm{Ga_{0.8}}\mathrm{As/AlAs}$
layers, while coupling between the cavities is mediated by a mirror
of 15.5 layer pairs of the same constitution. The coupling leads to
a \emph{cavity} mode splitting of 8\,meV, which was determined from
white-light reflection measurements. When the cavity modes are brought
into resonance with the quantum well excitons, four polariton branches
form. From white-light spectra, and with the help of the standard
transfer matrix approach, we deduced a Rabi splitting of about 5\,meV.
The quantum well exciton resonance is at 1607.7\,meV.

The particular scattering process in which we are interested is shown
on the right hand side of Fig.\,\ref{fig:doublecavity}: the system
is pumped on either of the two lower branches (energetically both
below the exciton line). The two resonant circles of the far-field
luminescence can be seen on the vertical projection, while the horizontal
projection shows the two branches as seen on a spectrometer (c.f.
Fig.\,\ref{fig:spectrum} below).

\begin{figure}[h]
\includegraphics[width=0.98\columnwidth]{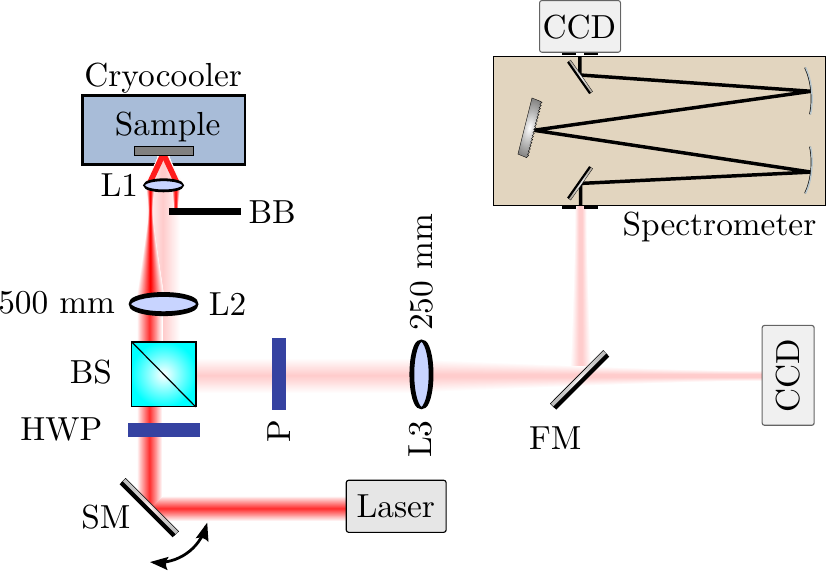}

\caption{Schematic drawing of the experimental setup. HWP: half-wave plate,
FM: flip mirror, SM: beam steering mirror, BS: beam splitter, P: polarizer,
L1, L2, L3: lens, BB: beam block.}

\label{fig:setup}
\end{figure}

Our experimental setup can be seen in Fig.\,\ref{fig:setup}. During
measurements, the sample was held at a temperature of 15\,K using
a Gifford-McMahon-type cryocooler. We would like to point out that
we could obtain qualitatively similar results even at 77\,K, but
the evaluation of data becomes difficult. First, excitons become ionized,
hence, polariton states are not very well defined, and second, the
phonon-induced background is much more dominant at those high temperatures.
The microcavity was resonantly excited by linearly polarized light
from a tunable Ti:Sapphire laser. In order to avoid non-linear effects,
intensities were kept at an average of 2\,$\mu\mathrm{W}$ (approximately
200\,$\mathrm{m}\mathrm{W}/\mathrm{cm}^{2}$), and we checked that
a further reduction in the power does not result in changes of the
\emph{relative} scattered intensities. An aspherical lens (L1) with
a numerical aperture of 0.82 was used for both focusing the laser
light onto the sample, and collecting emission. On the sample, we
directly measured a focal spot size of about 30\,$\mu$m with a Gaussian
intensity distribution. Far-field imaging was carried out by means
of a commercial CCD camera and a lens system consisting of two lenses
with focal lengths 500 (L2), and 200\,mm (L3), respectively. An imaging
spectrometer in combination with a second CCD was used for spectral
analysis. For better contrast, the reflected pump beam was blocked
by a small metal stripe (BB) inserted in the back focal plane of the
aspherical lens. In addition, we detected the emission from the sample
in the linear polarization basis cross-polarized to the pump. 

\section{Results and discussion}

\begin{figure}[h]
\includegraphics[width=1\columnwidth]{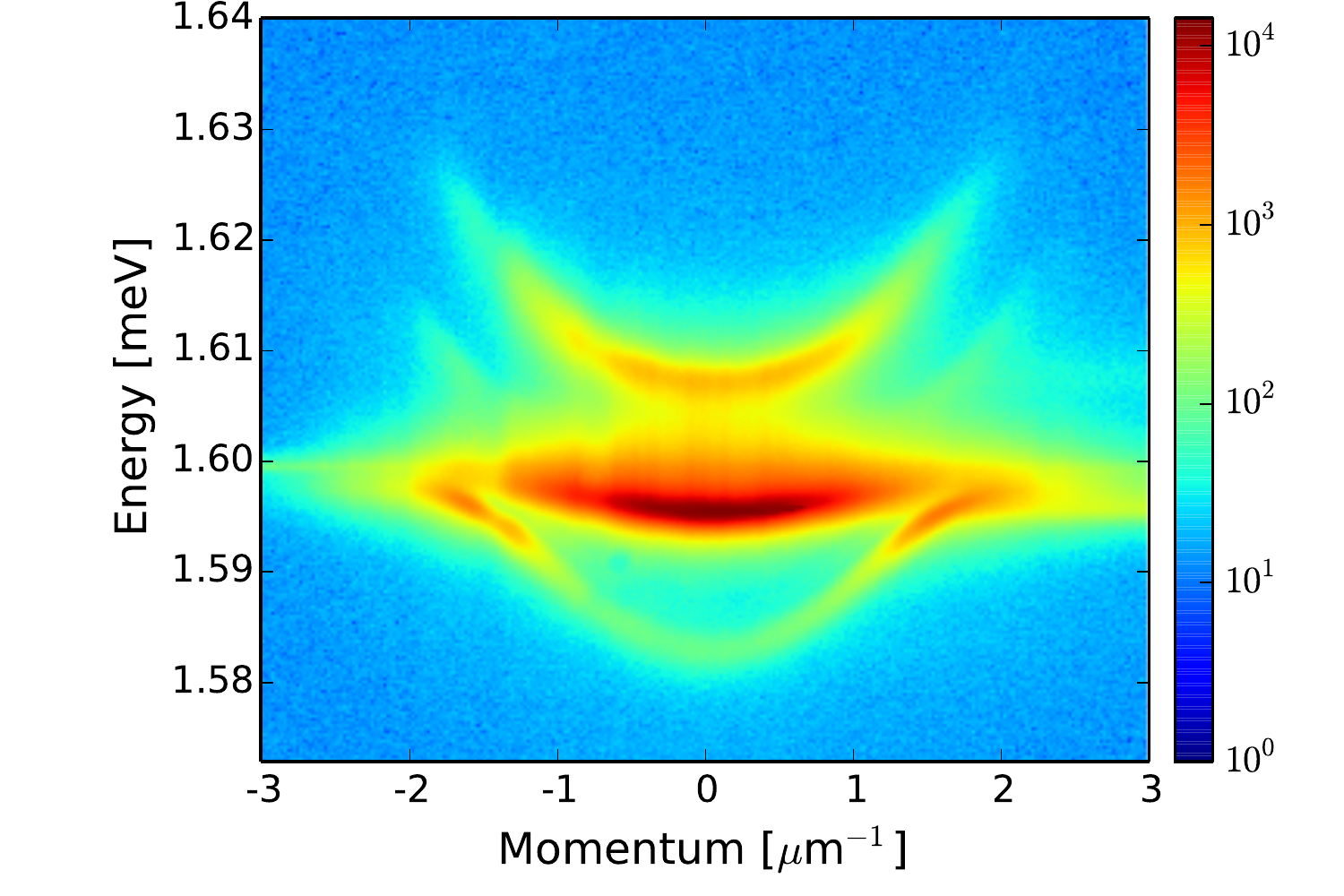}\caption{Spectrum of the polariton luminescence when the sample is excited
by a He-Ne laser.}

\label{fig:spectrum}
\end{figure}

A typical dispersion with the four polariton branches can be seen
in Fig.\,\ref{fig:spectrum}. In this case, the sample was probed
by off-resonant excitation at 633\,nm, and we collected the luminescence.
Note that the third polariton branch is hardly visible. 

\begin{figure}[h]
\includegraphics[width=0.98\columnwidth]{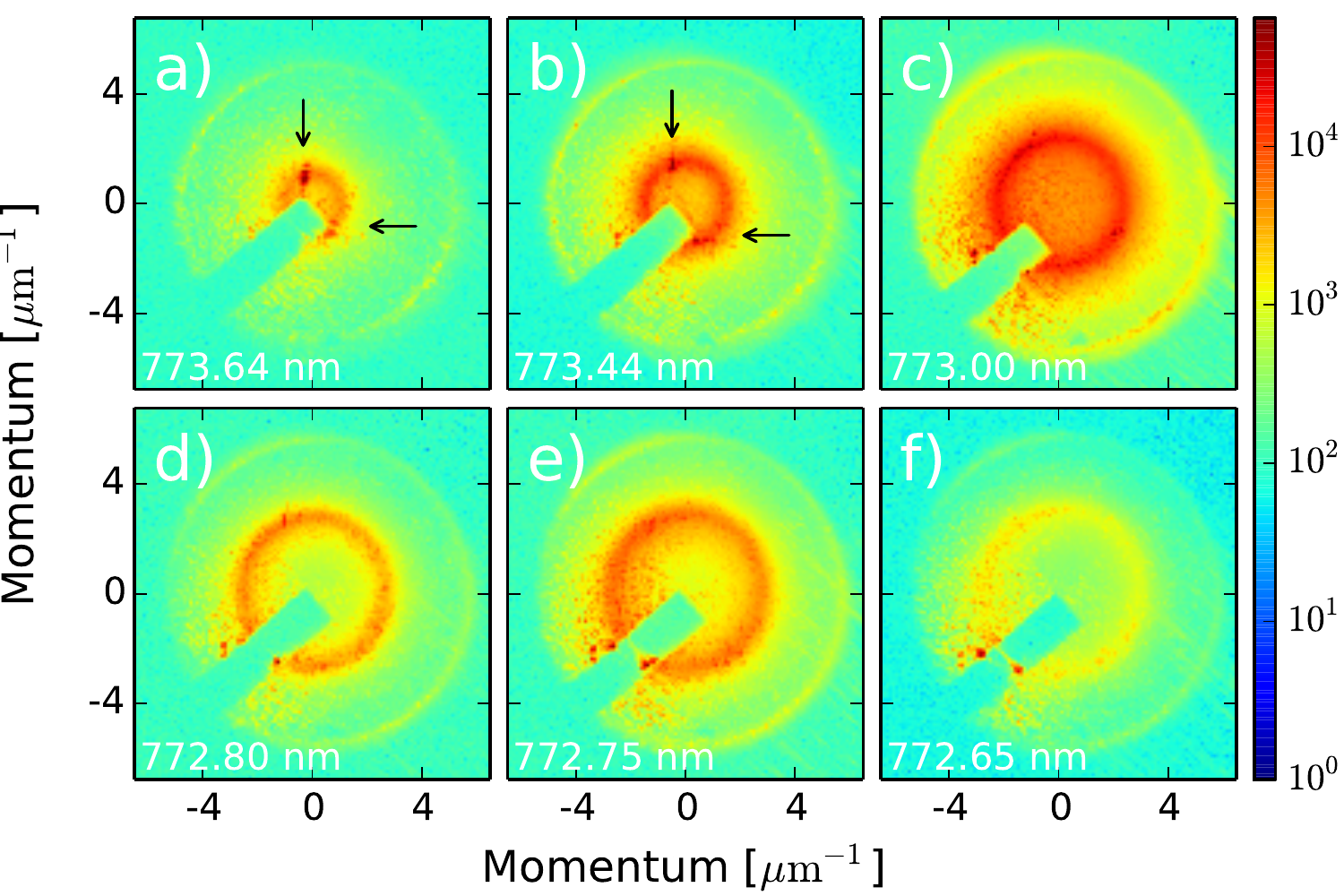}

\caption{Far-field images of the polariton emission when the sample is pumped
on the inner circle, i.e., second (anti-symmetric) branch. The laser
wavelength is given for each frame. The small arrows highlight the
preferential scattering directions of the cross-hatched dislocations.
The rectangular shadow is the small metal stripe covering the direct
laser reflection. }

\label{fig:far-field-experiment-second-branch}
\end{figure}

\begin{figure}[h]
\includegraphics[width=0.98\columnwidth]{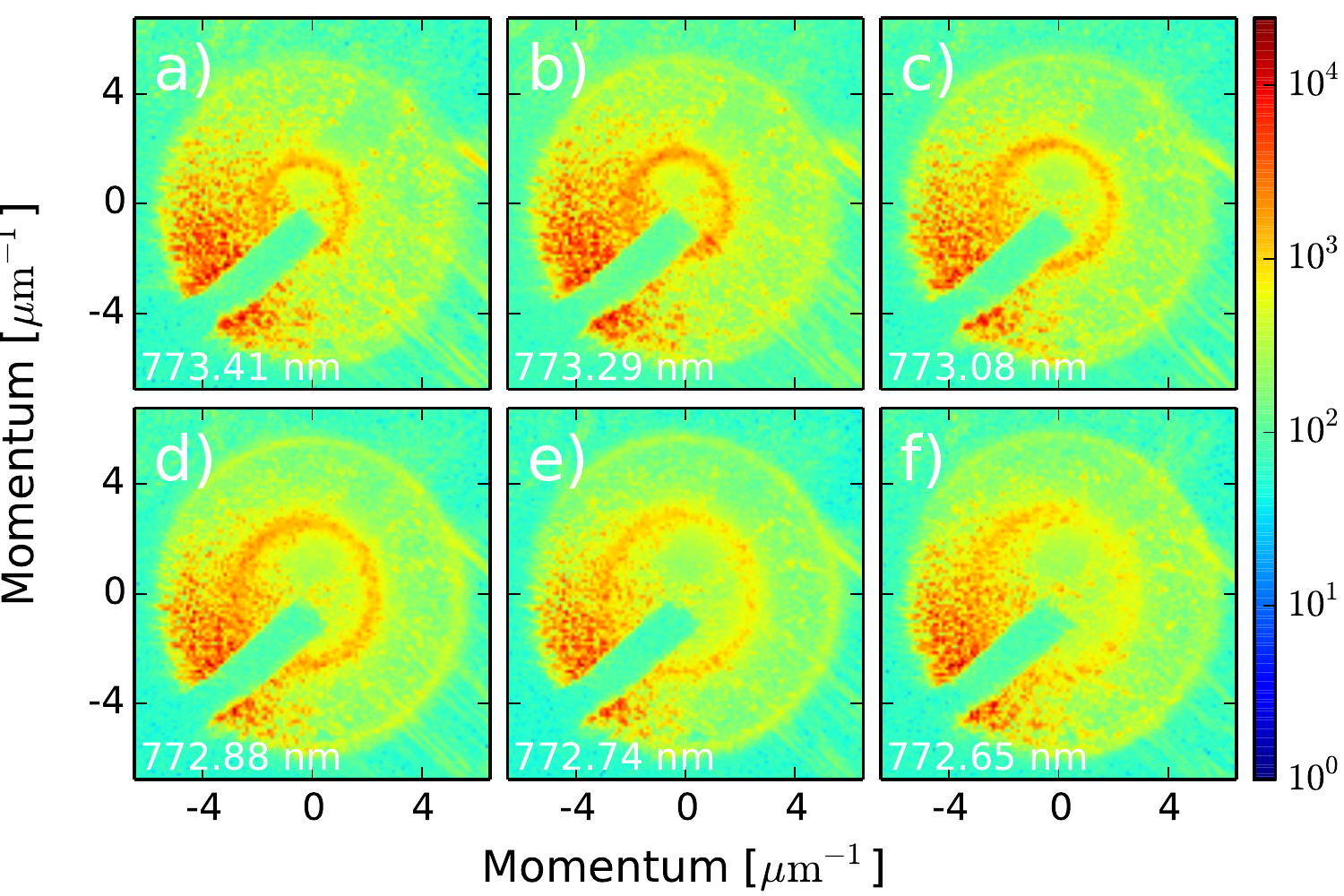}

\caption{Far-field images of the polariton emission when the sample is pumped
on the outer circle, i.e., the first (symmetric) branch. The laser
wavelength is given for each frame. The rectangular shadow is the
small wrench covering the direct laser reflection.}

\label{fig:far-field-experiment-first-branch}
\end{figure}

Exemplary RRS far field images are shown in Figs.\,\ref{fig:far-field-experiment-second-branch}-\ref{fig:far-field-experiment-first-branch}.
In these plots, the excitation position was fixed at a point, where
the cavity detuning is -10.1\,meV, and we changed the excitation
energy. The momentum was adjusted in a way that the laser light was
always resonant with one of the polariton branches. In the first series
(Fig.\,\ref{fig:far-field-experiment-second-branch}) the excitation
was on the second branch (inner circle). 

The resonant scattering gives rise to two rings, whose radii are dictated
by the local dispersion relations. The first ring results from scattering
into polariton states on the branch that is pumped, and it is equivalent
to the usual Rayleigh scattering in single cavities: owing to the
isotropic dispersion, the process conserves the absolute value of
the momentum. The second ring is generated by inter-branch scattering,
and is peculiar in the sense that the absolute value of the momentum
is no longer conserved. In Fig.\,\ref{fig:far-field-experiment-second-branch}(a-b),
one can also recognize a vertical and horizontal line intersecting
at the position of the pump momentum. Given that they are parallel
to the crystal axes, these lines can be attributed to the cross-hatch
dislocation mentioned earlier \cite{Zajac2012c}. 

Pumping with the same power on the first branch (outer ring, Fig.\,\ref{fig:far-field-experiment-first-branch})
leads to a similar emission pattern where again the two RRS rings
are visible, as shown in Fig.\,\ref{fig:far-field-experiment-first-branch}.
We should note, however, that the overall intensities are reduced
compared to Fig.\,\ref{fig:far-field-experiment-second-branch}.

The interpretation of inter-branch scattering observed in our samples
is that in the presence of disorder, momentum is not a good quantum
number, and, in the first approximation, the scattering process can
be described by Fermi's golden rule, which contains a constraint on
the energy for the transition probabilities. It is interesting to
note that without disorder, the two branches have even/odd parity
caused by the reflection symmetry of the cavities, and inter-branch
scattering would be forbidden by the symmetry. Therefore, disorder
is responsible for breaking not only the translational, but also the
reflection symmetry of the system. It can be shown that, if the disorder
potential in the two cavities is correlated/anti-correlated, the intra/inter-branch
scattering is suppressed. A thorough theoretical treatment of this
problem is given elsewhere \cite{Voros2014}.

\begin{figure}[h]
\includegraphics[width=0.98\columnwidth]{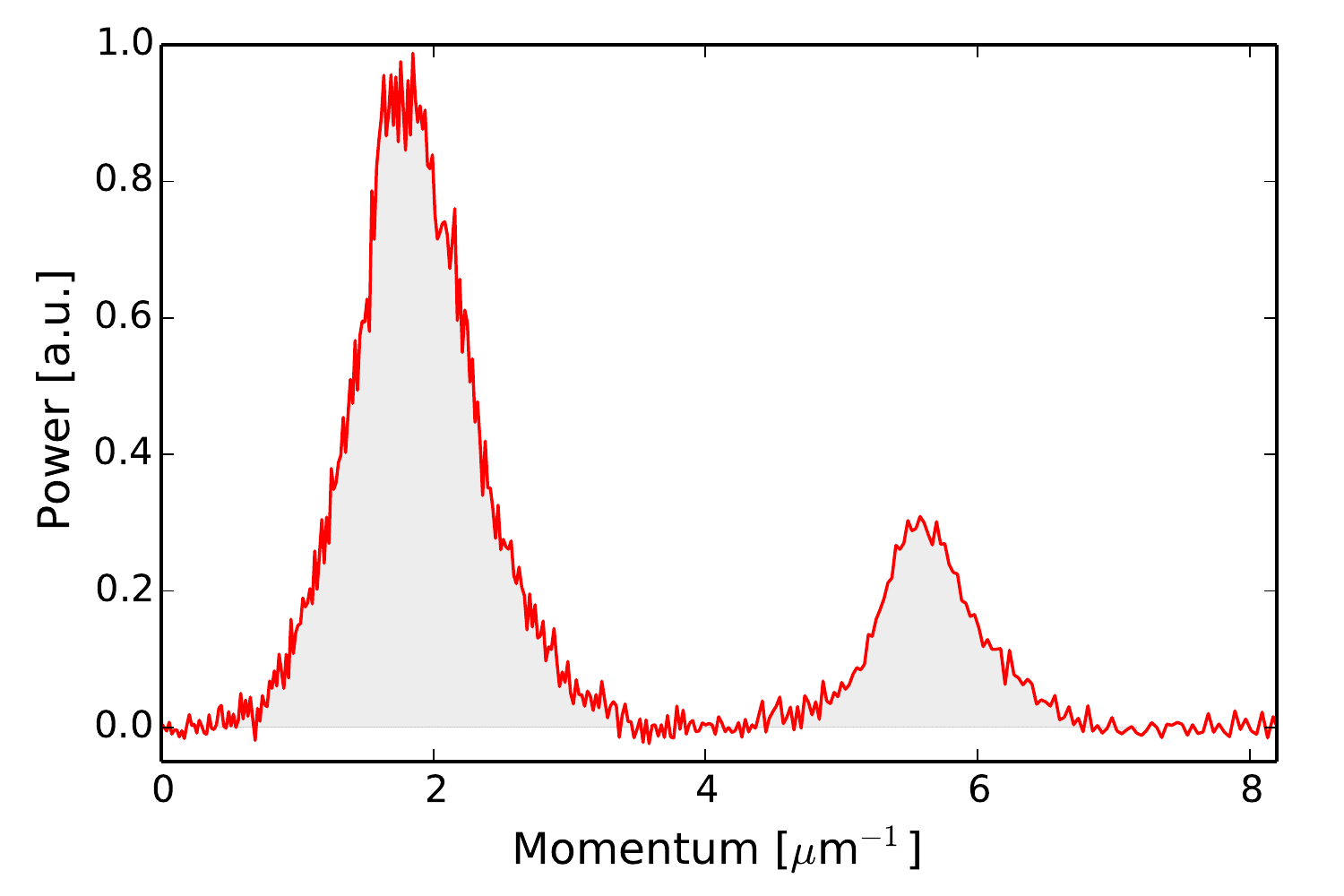}

\caption{Azimuthally integrated Rayleigh-scattered power. The total powers
were calculated by integrating over the two shaded domains. The maximum
of the power was normalized to 1.}

\label{fig:radially_integrated}
\end{figure}

These results can be made more quantitative by comparing the luminescence
powers of the two rings. As long as non-linear effects can be ruled
out (low excitation power), the relative intensities are independent
of the pump power, and are only a function of the photonic disorder
potentials in the two cavities. In Fig.\,\ref{fig:radially_integrated},
we show the azimuthally integrated power as a function of the absolute
value of the momentum. The origin of the polar coordinate system was
determined by first visually fitting a circle on the inner ring. However,
as long as the two rings are well-separated (which is the case for
all our measurements), the exact position of the origin is not crucial:
the measured intensities for each ring will be the same, only the
profiles change slightly. The intensities were normalized to a maximum
of 1. The two peaks corresponding to the two Rayleigh rings are clearly
identifiable. The shaded areas denote the range over which we integrated
the power to measure the total scattering rate in subsequent figures.

\begin{figure}[h]
\includegraphics[width=0.98\columnwidth]{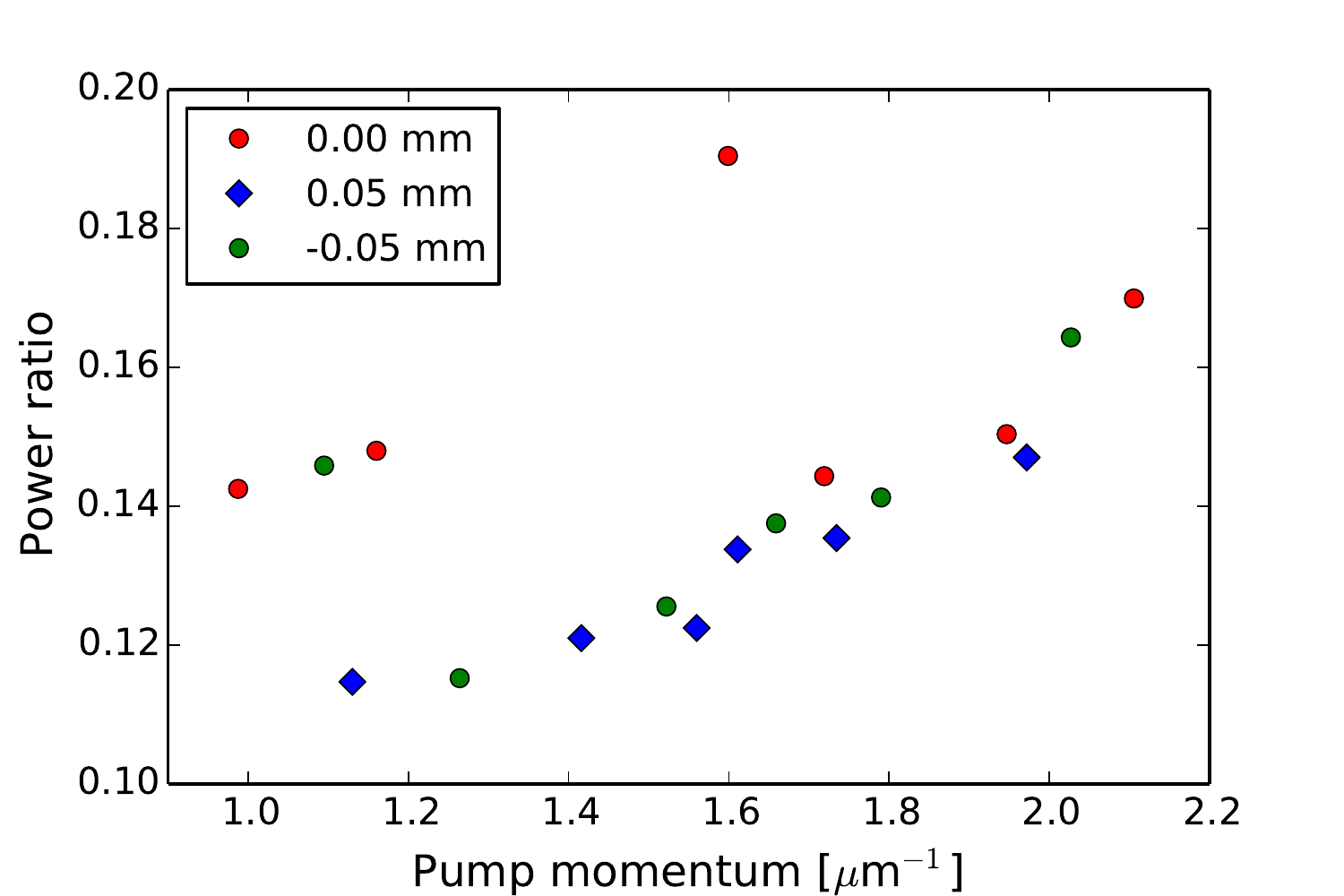}

\caption{Relative scattering power as a function of the excitation momentum
at three different measurements points, taken 0.05\,mm apart. The
statistical errorbars are too small to be visible.}

\label{fig:scattering_ratio_vs_momentum}
\end{figure}

In Fig.\,\ref{fig:scattering_ratio_vs_momentum} we plot the ratio
of the powers for three positions on the sample as a function of the
pump momentum. The detuning was the same for all three points. The
power ratio was defined as the total power of the outer ring divided
by the total power of the inner ring, as shown in Fig.\,\ref{fig:radially_integrated}.
In this case, we excited on the inner ring (second branch). The laser
wavelength was adjusted to match the polariton dispersion for each
momentum, and we recorded 5 images for each case. The displayed data
points are the average of the 5 measurements. The three curves run
close to each other, and they obey a general slightly increasing trend
as the momentum increases, but there are some local variations. 

\begin{figure}
\includegraphics[width=0.98\columnwidth]{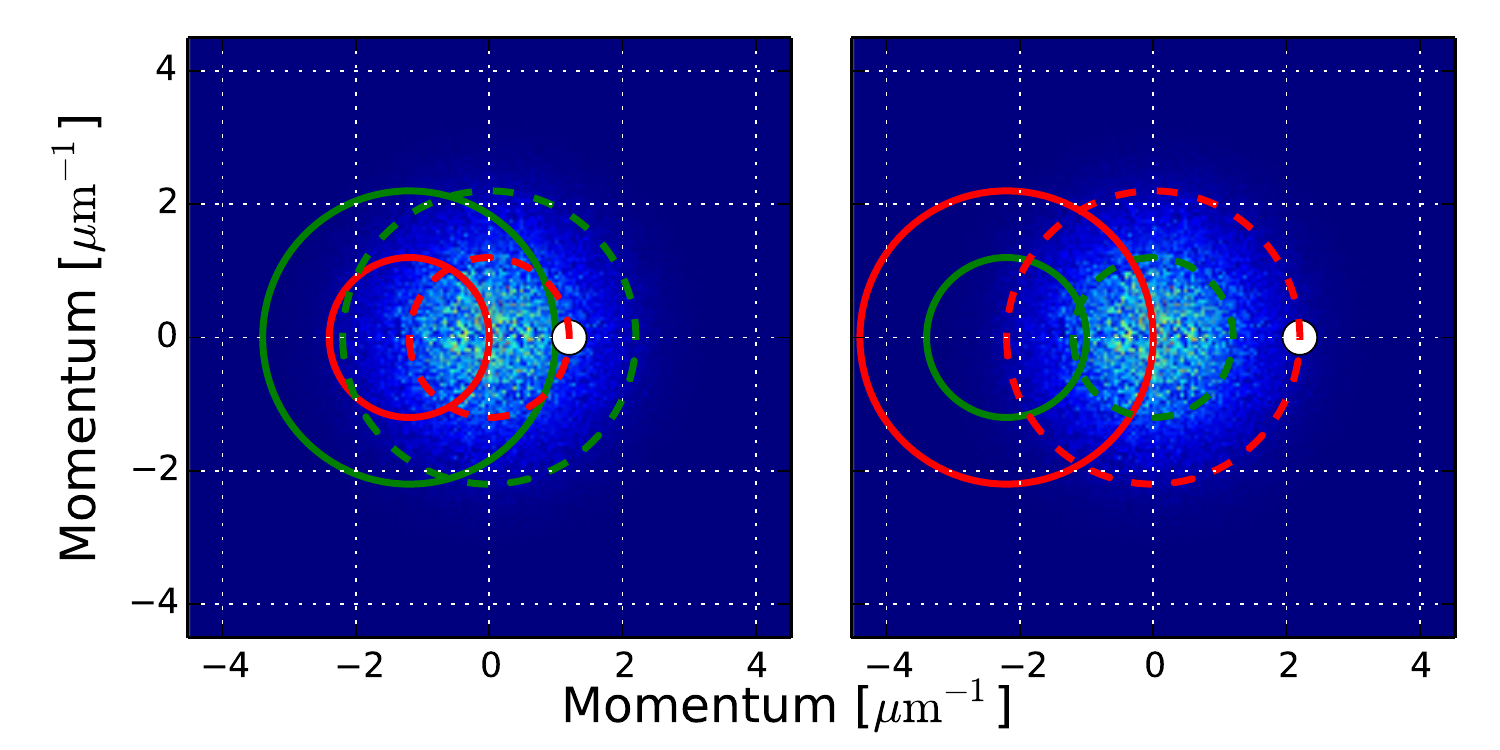}

\caption{Schematic drawing of the scattering wave vectors. The pump laser is
represented by the white circle, and scattering occurs on the dashed
circles. The solid lines are the difference wave vectors. The noisy
background is the Fourier transform of the disorder potential, in
this case assumed to have Gaussian correlations with a correlation
length of $0.5\,\mu$m.}

\label{fig:coupled_scattering}
\end{figure}

The physical reason for a change in the scattering amplitudes is the
following. By Fermi's golden rule, the transition probabilities are
proportional to the square of the matrix elements taken at the difference
wave vector. The matrix elements themselves are the Fourier transforms
of the scattering potential. Now, for technical reasons, the disorder
potential usually has some \emph{transverse }correlation whose Fourier
transform displays a maximum at zero momentum. The transition probabilities,
and hence, the power of the Rayleigh rings can be obtained by integrating
this Fourier transform over all allowed wave vectors, i.e., over the
circle of difference wave vectors as shown in Fig.\,\ref{fig:coupled_scattering}.
When the pump momentum changes, so does the radius of the scattering
ring, and thus, a different domain of the Fourier transform will be
sampled. This leads to a different intensity distribution, and different
total power.

\begin{figure}[h]
\includegraphics[width=0.98\columnwidth]{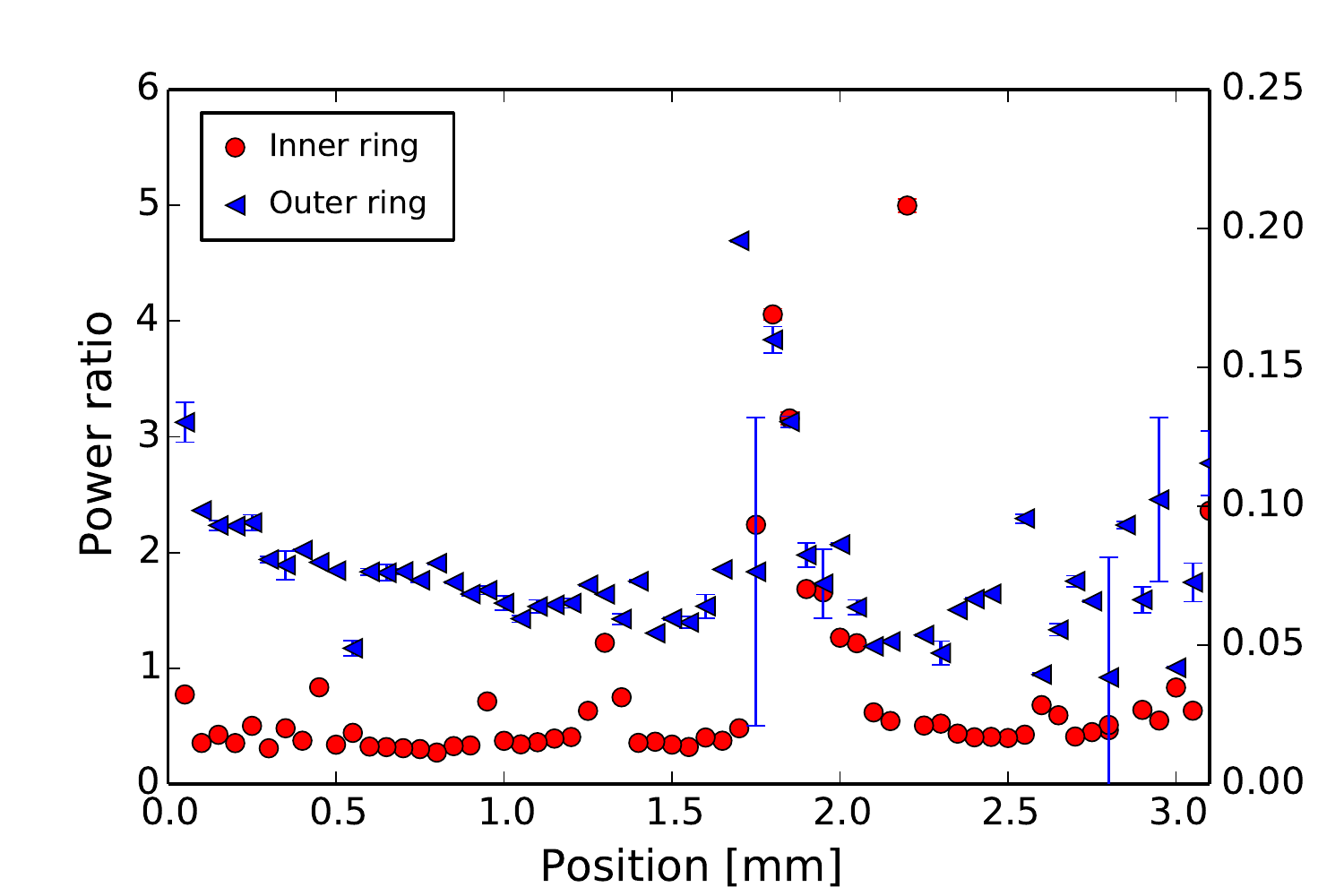}

\caption{The power ratio as a function of the spatial coordinate. Outer ring
pumping is plotted on the right, while inner ring pumping is shown
on the left axis. }

\label{fig:innerfine}
\end{figure}

In Fig.\,\ref{fig:innerfine}, we plot the dependence of the relative
power on the spatial coordinate for two cases, once when the sample
is pumped on the inner ring, and once on the outer ring. The measurement
points were chosen along a line perpendicular to the wafer's gradient,
so as to keep the detuning constant. The first measurement position
was taken as the reference point of the coordinate system. 

The two curves display similar behavior: apart from small variations,
and a slight and smooth overall change, there is a well-defined maximum
at position 1.75\,mm. It is also worth noting that the outer ring,
as in Figs.\,\ref{fig:far-field-experiment-second-branch}-\ref{fig:far-field-experiment-first-branch},
is generally dimmer (the power ratio is smaller than 1) than the inner
ring. 

It can be shown that the amount of inter-branch scattering is related
to the inter-cavity correlations of the disorder potential \cite{Voros2014}.
One might ask, why these correlations would develop in the first place,
if the disorder is truly random. The transverse correlations are related
to the spatial distribution of material beams in the epitaxy equipment.
It can be expected that, if the beam is not completely uniform in
space, then any irregularities will vanish over a finite distance
\cite{Savona2007}. By the same token, since the cavities are grown
sequentially in time, the longitudinal correlations reveal information
about the long-term stability of the deposition process. Again, it
is reasonable to expect that some deviations from the average deposition
rate will be sustained over longer periods of time, simply because
they are related to long-term parameters (e.g., geometry, pressure/temperature
distribution, target positions) of the fabrication system. This also
implies that the disorder in the two cavities will probably be correlated,
and not anti-correlated.

\begin{figure}[h]
\includegraphics[width=0.98\columnwidth]{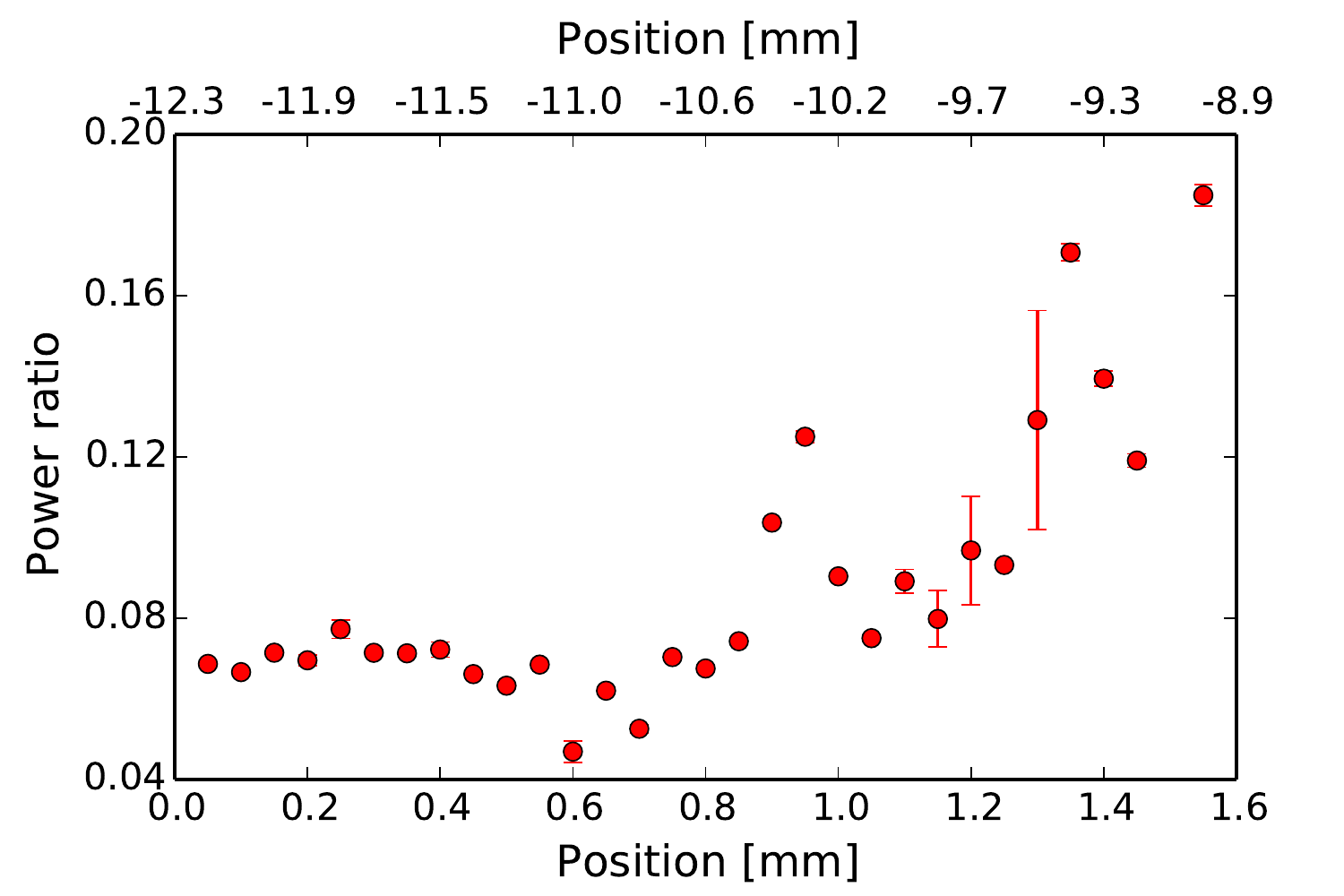}

\caption{Inter-branch scattering power as a function of detuning, when the
excitation point is scanned along the sample gradient. Positions are
taken from the first measurement point as reference. Zero is closer
to the wafer's center. The excitation wavelength was 773.13 nm for
all points. }

\label{fig:mar01_detscan2}
\end{figure}

In Fig.\,\ref{fig:mar01_detscan2}, we plot the inter-branch scattering
power as a function of the detuning. (We defined detuning as the energy
difference between the exciton, and the uncoupled photon state. This
latter one is the mean of the coupled photon energies.) In this case,
the excitation spot was shifted along the wafer's gradient, while
the excitation energy was kept constant at 1603.66\,meV, and the
excitation momentum was always adjusted to match the polariton resonance.
In principle, the detuning should not influence the strength of inter-branch
scattering. Since RRS is caused by photonic disorder, its strength
will be proportional to the photon content of the particular polariton
state. However, this depends on the Hopfield coefficients, which are
simply a function of the energy difference between the state in question,
and the exciton's energy \cite{Einkemmer2013}. Therefore, for resonant
phenomena, only the absolute scattering amplitude depends on the energy,
but not the relative powers. For this reason, any change in the relative
intensities should be regarded as a change in the longitudinal correlations,
and not as an effect of the detuning.\textcolor{blue}{{} }In Fig.\,\ref{fig:mar01_detscan2},
we recognize an increase in inter-branch scattering for smaller red
detuning, which also happen to lie farther from the center of the
wafer. While the trend may not be significant, it could indicate increasing
disorder correlation.

\section{Conclusion}

In conclusion, we have presented an experimental study of polariton
Rayleigh scattering in coupled microcavities. We have shown that the
photonic disorder that breaks both the translational and reflection
symmetry of the system re-distributes the polaritons to two branches,
and this results in two concentric emission rings in the far field.
The absolute value of the initial momentum is conserved only on one
of the branches. We also argued that by inspecting the relative scattered
intensities on the two rings, one can infer the long-term stability
of the deposition process. 
\begin{acknowledgments}
The authors gratefully acknowledge financial support from the Austrian
Science Fund, FWF, project number P-22979-N16.
\end{acknowledgments}
\bibliographystyle{apsrev4-1}
\bibliography{/home/v923z/papers/polariton/polariton}

\end{document}